\documentclass[12pt]{iopart}
\usepackage{epsfig,graphicx,psfrag,amssymb,amsfonts,latexsym,color,dcolumn,bm}

\begin{document}
\title{Quantum open systems approach to the dynamical Casimir effect}

\author{F.C. Lombardo and  F. D. Mazzitelli}

\address{Departamento de Fisica J.J. Giambiagi, Facultad de Ciencias Exactas y Naturales,
Universidad de Buenos Aires, Ciudad Universitaria, Pabell\'on 1,
1428 Buenos Aires, Argentina}

\ead{lombardo@df.uba.ar, fmazzi@df.uba.ar}

\begin{abstract}
We analyze the introduction of dissipative effects in the study of the dynamical Casimir effect. We consider a toy model for an electromagnetic cavity that contains a semiconducting thin shell, which
is irradiated with short laser pulses in order to produce periodic oscillations of its conductivity. The coupling between the quantum field in the cavity  and the microscopic degrees of freedom of the shell induces dissipation and noise in the dynamics of the field. 
We argue that the photon creation process should be described in terms of a damped oscillator with nonlocal dissipation and 
colored noise.

\end{abstract}




\def\ii{\'{\i}}
\def\beq{\begin{equation}}
\def\eeq{\end{equation}}
\def\lp{\left(}
\def\rp{\right)}
\def\lb{\left[}
\def\rb{\right]}
\def\om{\omega}
\def\La{\Lambda}
\def\la{\lambda}
\def\ck{\chi_k}

\section{Introduction}
The motion of a neutral body  distorts the quantum electromagnetic vacuum and may  induce
photon creation. This fact was first pointed out many years ago \cite{Moore}. 
The radiation of  a single mirror is, in general, an extremely small effect. For this reason, during the seventies, 
the radiation of moving mirrors was mainly analyzed because its formal relation with black hole evaporation
\cite{Fulling}. However, it was later 
pointed out that, in  a resonant cavity, photon creation could be enhanced by secular effects: if the length of the cavity oscillates with a frequency which is twice the frequency of an eigenmode of the static cavity, resonant effects may produce a number of photons that, under ideal conditions, grows exponentially with time \cite{cavities}. 

Although intensively studied from the theoretical point of view, this ``motion induced radiation'' 
is extremely difficult to measure. On the one hand, cavities with a very high $Q$ are needed in order to have
an exponential growth during a significative amount of time. On the other hand, and more importantly, for microwave cavities the 
mirror should oscillate at extremely high frequencies, of the order of Ghz. This is a challenge
from an experimental point of view, although there are feasible proposals based on the use of 
nanoresonators in order to accomplish it \cite{Onofrio}.

Photon creation can occur whenever there are time dependent external conditions: moving boundaries and/or time dependent electromagnetic properties. These broad class of phenomena is usually denominated "Dynamical Casimir Effect" (DCE) (for a complete list of publications see \cite{reviews}).  In particular, a setup that has attracted both theoretical and experimental attention is the possibility of using short laser pulses in order to produce periodic variations of the conductivity of a semiconductor layer placed inside a microwave cavity. The fast changes in the conductivity induce a periodic variation in the effective length of the cavity, and 
therefore the creation of photon pairs \cite{Lozovik}. This setup has been analyzed at the theoretical level \cite{Croccemir,Dodonovmir}, 
and there is an ongoing experiment aimed at the detection of the motion induced
radiation \cite{Padova}.  More recently, the possibility of changing the effective length of 
a superconducting coplanar waveguide terminated by a SQUID using a time dependent magnetic field has also been put forward \cite{Nori}.

In this paper we will be mainly concerned with the irradiated-semiconductor setup. It has been 
pointed out that dissipative effects may play an important role in this experiment, that could induce a    
significant deviation from the expected exponential growth in the number of created photons
\cite{Dodonovmir}. To our knowledge, up to now there is no theoretical model that includes, from first principles, 
effects of dissipation and noise. There are, on the other hand, phenomenological models \cite{Dodonovdis} to estimate these effects. The idea is the following: in a resonant situation, if the spectrum of the cavity is not equidistant,
the photons are created in a single mode. The dynamics of this mode can be described, in the absence of dissipation, in terms of a harmonic oscillator with time dependent frequency. A large number of photons is created due to  parametric resonance. A phenomenological way of introducing dissipation and noise is to describe the temporal evolution of the modes in terms of a damped oscillator. Mathematical consistency for quantization requires the introduction of an additional stochastic force,
the noise. Physically, this is nothing but the well known fluctuation-dissipation relation.

The aim of this paper is to provide a first step towards a description of the DCE from first principles, including dissipation and noise. The natural  arena for this description is the theory of quantum
open systems. We will consider the electromagnetic field as our "system", while the degrees of freedom
on the semiconductor  will be the "environment". Under very general assumptions, is it possible 
to show that the dynamics of the electromagnetic field will be described by a Langevin equation (Section 3). In general, this equation will include a nonlocal dissipative term and a colored noise, both
concentrated on the position of the layer. We will use this Langevin equation to describe 
the dynamics of the resonant mode (Section 4). We will see that the 
evolution of a single mode is in general more complex than that of a damped oscillator with white 
noise, and that all the effects come from the non trivial boundary conditions at the position of the slab. 

In order to avoid unnecessary complications  we will consider several simplifying assumptions. On the one hand we will 
consider a quantum scalar field in $1+1$ dimensions instead of the electromagnetic field in a $3+1$ cavity. On the other hand, 
we will assume that this quantum field is linearly coupled to the degrees of freedom of the layer. Finally, we will consider a 
very thin semiconductor layer. Even with these simplifications, we hope the resulting model to  
contain the main physics of the problem. This model can be considered as a 
generalization of the theory described in Ref.\cite{Croccemir}, reviewed in Section 2, to the dissipative case.

\section{A model without dissipation}

In this section we will review the model of Ref. \cite{Croccemir}, which we will use as starting 
point to include dissipative effects in the rest of the paper.

We consider a massless scalar field 
within a cavity with perfect conducting walls. For simplicity,  
we consider the 1+1 dimensional case, where we consider a cavity of size $L$. At the mid point of the cavity
($x=L/2$) a thin film of semiconducting material is located. We
model the conductivity properties  of such material by a potential
$V(t)$: The ideal limit of perfect conductivity corresponds to $V
\rightarrow \infty$, and $V \rightarrow 0$ to a `transparent'
material. This potential varies
between a minimum value, $V_0$, and a maximum $V_{\rm max}$. The
Lagrangian of the scalar field within the cavity is given by 
\beq
{\mathcal L}=\frac{1}{2} \partial_{\mu}\phi\partial^{\mu}\phi -
\frac{V(t)}{2} \delta(x-L/2) \phi^2, \label{themodel} 
\eeq 
where
$\delta(x)$ is the one-dimensional Dirac delta function.
The use of an infinitely thin film is justified as long as the width of the slab is much 
smaller than 
the wavelengths of the relevant  modes in the cavity. 
The
corresponding Lagrange equation reads, \beq (\partial_x^2 -
\partial_t^2) \phi = V(t) \delta(x-L/2)\phi.
\label{fieldequation} \eeq
As shown in Ref.\cite{barton}, this model describes
plane-polarized
electromagnetic field propagating normally to an infinitesimally
thin jellium-type plasma sheet. The strength of the potential 
is given by 
\begin{equation}
V=4 \pi n_s e^2/m\, ,
\label{plasma}
\end{equation}
where $n_s$
is the surface charge density of free carriers in the
sheet, $e$ is the electron charge, and $m$ is the
effective mass of the free carriers.
When a laser field suddenly
impinges on the plasma sheet, it produces time-dependent changes in the
surface charge density $n_s(t)$ that induce a time variation in
the conductivity $V(t)$.

For the sake of clarity we divide the cavity into two regions: region I ($0\leq x \leq L/2$)
and region II ($L/2 \leq x \leq L$). Perfect conductivity at the edges of the cavity
imposes the following boundary conditions
\begin{equation} \phi_{\rm I}(x=0,t)=\phi_{\rm II}(x=L,t)=0. 
\label{BC1}
\end{equation}
The presence of the semiconducting film introduces a discontinuity in the spatial derivative
of the field, while the field itself remains continuous,
\begin{eqnarray}
&&\phi_{\rm I}(x=L/2,t)=\phi_{\rm II}(x=L/2,t), \nonumber \\
&&\partial_x\phi_{\rm I}(x=L/2,t)-\partial_x \phi_{\rm II}(x=L/2,t)=-V(t) \phi(x=L/2,t).
\label{BC2}
\end{eqnarray}
This can be seen by integrating out the field equation
(Eq. (\ref{fieldequation})) in the neighborhood of the film.

We consider the set of functions 
\begin{equation}
\psi_{m}({x},t) = \sqrt{\frac{2}{L}}
\sin\left(k_{m}(t)\,x\right)\, , \label{sol2}
\end{equation}
where $k_{m}(t)$ is the $m$-th positive solution to the following
transcendental equation \beq 2 k_{m}
=- V(t)\tan\left(\frac{k_{m}L}{2}\right) .
\label{trascendentalequation} \eeq 
Note that $\psi_{m}$ depends on $t$ through $k_{m}(t)$.

Let us define
\begin{equation}
\Psi_{m}({x},t) = \left\{   \begin{array}{ll}
                 \psi_{m}(x,t) & 0 \le x  \le L/2 \\
                 -\psi_{m}(x-L,t) & L/2 \le x \le L
                      \end{array}
        \right.
\end{equation}
These functions satisfy the boundary conditions Eq.(\ref{BC1}) and
Eq.(\ref{BC2}), and the orthogonality relations $\left(\Psi_{m}, \Psi_{n}
\right)  = \left[ 1 - \sin(k_m(t) L) / k_m(t) L \right]
\delta_{m,n}$, where we have used the usual inner product in
the interval $[0,L]$. 

For $t\leq 0$ the slab is not irradiated, consequently $V$ is
independent of time and has the value $V_0$. The modes of the
quantum scalar field that satisfy the Klein Gordon equation
(\ref{fieldequation}) are
\begin{equation}
u_{m}({x},t)=\frac{e^{-i \bar \omega_{m} t}}{ \sqrt{
2\bar\omega_{m}}}\Psi_{m}({x},0)\,\,\, , \label{u}
\end{equation}
where $\bar\omega_{m}=k_m^0$,
 and $k_m^0$ is the $m$-th
solution to Eq.(\ref{trascendentalequation}) for $V=V_0$. At
$t=0$ the potential starts to change in time and the set $\{ k_m
\}$ of the eigenfrequencies of the cavity acquires a time
dependence through Eq.(\ref{trascendentalequation}).

We expand the field operator $\phi$ as
\begin{equation}
\phi({x},t) =
\sum_{m} \left[b_{m} u_{m}({x}, t)
+ b^\dagger_{m} u^*_{m}({x}, t) \right],
\end{equation}
where $b_{m}$ and $b^\dagger_{m}$  are annihilation and creation operators, respectively.
There is another set of solutions of the Klein Gordon equation that
satisfy all boundary conditions and have a 
node at $x=L/2$. For this reason their dynamics is not
affected by the presence of the slab. Moreover, these modes are decoupled from the modes $u_{m}$ 
thanks to the orthogonality 
conditions.
Therefore, we will only consider the evolution of the modes given in Eq.(\ref{u}).

For $t\geq 0$ we write the expansion of the field mode $u_{s}$ as
\beq
u_{s}({x},t>0) = \sum_{m} P_{m}^{({
s})}(t) \Psi_{m}({x},t) .
\label{singlefunction}
\eeq
Replacing this expression into $(\partial_x^2 -
\partial^2_t) u_{s}=0$  we find
\begin{equation}
\ddot{P}_{n}^{({s})} +  k_n^2(t) P_{
n}^{({s})} = - \sum_{m} \left[ \left( 2 \dot{P}_{
m}^{({s})} \dot{k}_{m} + P_{m}^{({s})}
\ddot{k}_{m} \right)   g_{m n}^{(A)} + P_{
m}^{({s})} \dot{k}_{m}^2  g_{m n}^{(B)} \right]\, ,
\label{eqP}
\end{equation}
where the coefficients
$g_{m n}^{(i)}$ read
\begin{eqnarray}
g_{m n}^{(A)} &=&  \frac{1}{\left(\Psi_{n}, \Psi_{n}
\right) }
\,   \left(  \frac{\partial \Psi_{m}}{\partial k_m} , \Psi_{n} \right) , \nonumber \\
g_{m n}^{(B)} &=&  \frac{1}{\left( \Psi_{n}, \Psi_{n}
\right)} \,  \left(  \frac{\partial^2 \Psi_{m }}{\partial k_m^2}
, \Psi_{n} \right). \label{coef12}
\end{eqnarray}
Note
that $\bar\omega_{m} =k_m(0)$. 

We are interested in the number of photons created inside the
cavity. Hence we focus in resonance effects induced by periodic
oscillations in the conductivity $V(t)$, which translates into
effective  periodic changes in the modes of the scalar
field. Therefore we start by considering a time dependent
conductivity given by
\begin{equation}
V(t) = V_0 + \left(V_{\rm max} - V_0\right) f(t)\,\,\, ,
\label{vgen}
\end{equation}
where $f(t)$ is a periodic and non negative function,
$f(t)=f(t+T)\geq 0$, that vanishes at $t=0$ and attains its maximum
at $f(\tau_e)=1$. In each period, $f(t)$ describes the excitation
and relaxation of the semiconductor produced by the laser pulse.
Typically, the characteristic time of excitation $\tau_e$ is the
smallest time scale and satisfies $\tau_e\ll T$. We expand $f(t)$
in a Fourier series
\begin{equation}
f(t) = f_0 + \sum_{j=1}^{\infty} f_j \cos(j \Omega t + c_j) ,
\label{expand}
\end{equation}
where $\Omega=2\pi/T$. Since $\tau_e$ is the smallest time scale,
on general grounds we expect the first $T/\tau_e$ terms in the
above series to be relevant. 

Under certain constraints, large changes in $V$  induce only small
variations in $k$ through the transcendental relation between $k$
and $V$ (see Eq.(\ref{trascendentalequation})). In this case, a
perturbative treatment is valid and a linearization of such
relation is appropriate. Accordingly we write 
\beq k_n(t)=k_n^0
(1+\epsilon_n f(t)) , \label{kdet} 
\eeq 
where $\epsilon_n$ is
obtained after replacing Eqs. (\ref{vgen},\ref{kdet}) into Eq.
(\ref{trascendentalequation}) and expanding it to first order in
$\epsilon_n$. The result is 
\beq \epsilon_n = \frac{V_{\rm
max}-V_0}{L (k_n^0)^2 + V_0 \left(1+\frac{V_0 L}{4}\right)}
.\label{epsilon} \eeq 
The restriction for the validity of the
perturbative treatment is $V_0 L \gg V_{\rm max} / V_0>1$. These
conditions are satisfied for realistic values of $L$, $V_0$, and
$V_{\rm max}$.  
Indeed, from Eq.(\ref{plasma}), and using known values for the
conductivities of good conductors, we can fix $V_{\rm max} =
10^{16} {\rm m}^{-1}$.  When the laser field is not applied  we
can set $V_0 = 10^{10} {\rm m}^{-1} - 10^{13} {\rm m}^{-1}$, the
range of values for different semiconductors. For a cavity of size
$L_x\simeq 10^{-2} {\rm m}$,  and when the ratio between the maximum and
minimum conductivities is in the range $10^3 \leq V_{\rm
max}/{V_0} \leq 10^6$, we obtain from Eq.(\ref{epsilon}) 
small values of $\epsilon$, $10^{-8} \leq \epsilon
\leq 10^{-2}$.
It is worth noticing that we are
interested in low eigenfrequencies, for which $k(t)\sim {\mathcal
O}(L^{-1})$. Nonetheless the perturbative treatment is also
valid for $k\sim {\mathcal O}(V)$.

In what follows we will only consider expressions to first order
in $\epsilon_n$. To analyze the
possibility of parametric resonance we write the time dependent frequency given in
Eq. (\ref{kdet}) as,
\beq k_n(t) = \tilde{k}^{0}_n (1+\epsilon_n
(f-f_0)),
\eeq
where $\tilde{k}^0_n \equiv k^0_n (1+\epsilon_{\rm
n}f_0)$ is a `renormalized' frequency. The equation for the
coefficients $P^{({s})}_{m}(t)$ (Eq. (\ref{eqP})) can now
we written to first order in $\epsilon_n$ as,
\begin{eqnarray}
\ddot{P}^{({s})}_{n} +  (\tilde k_n^0)^2 P^{({s})}_{n} &=& -2
\epsilon_n (k_n^0)^2 (f-f_0) P^{({s})}_{n} \nonumber \\ &-& \sum_{m}
\left[ 2 \dot{P}_{m}^{({s})}  \epsilon_m k^0_m  \dot f + \
P_{m}^{({s})}  \epsilon_m  k^0_m \ddot f \right]  g_{m
n}^{(A)} +  {\mathcal O}(\epsilon^2).
\end{eqnarray}
This equation describes a set of coupled
harmonic oscillators with periodic frequencies and couplings. It
is of the same form as the equations that describe the modes of a
scalar field in a three dimensional cavity with an oscillating
boundary. A naive perturbative solution of previous equations
in powers of $\epsilon_n$ breaks down after a short amount of time (this happens
for particular values of the external frequency such that there is a resonant
coupling with eigenfrequencies of the cavity). In order to analyze the long time
behavior of the solutions the equations can be solved  using  multiple scale analysis (MSA)
\cite{Crocce3d}. 

As shown in detail in Ref.\cite{Croccemir}, if the spectrum of the cavity is not equidistant, then the photons are 
created mainly in the resonant mode. This means
that this particular mode decouples from the rest, and the subsequent quantum dynamics
can be described starting from the equation
 \beq
\ddot{P}^{({
n})}_{n} + (\tilde k_n^0)^2  P^{({n})}_{n} + 2
\epsilon_n (k_n^0)^2 (f-f_0) P^{({n})}_{n} =0\, ,\eeq
that is, a harmonic oscillator with time dependent frequency. When $\Omega= 2\tilde k_n^0$,  the number of created 
photons grows exponentially due to parametric resonance.

\section{Quantum open systems: The Langevin equation for the field}

In the previous section, the starting point was the action given in Eq.(\ref{themodel}), which describes a 
quantum field with a time-dependent mass term concentrated on the position of the slab. In a more realistic model, the massless scalar field should be coupled to the microscopic degrees of freedom 
on the slab, and its  dynamics will be described in terms of the effective action that results
after integrating the microscopic degrees of freedom.

In static situations, it is enough to compute the in-out or the Euclidean effective action, and to read the zero point energy of the system from the generating functional. However, if we are interested in the 
temporal evolution of the quantum field, it is necessary to compute the in-in or CTP effective action \cite{CTP}. In this 
direction, we will consider the scalar field as a quantum open system, which is coupled to an 
external environment composed by the internal degrees of freedom in the slab. Therefore, we will  
consider model composed by a classical action given by 

\begin{equation}
S[\phi,q_n] = S_0[\phi] + S_0[q_n] + S_{\rm int}[\phi, q_n], 
\end{equation}
where $S_0[\phi]$ and $S_0[q_n]$ are the free actions for the massless scalar field 
and for the infinite set of harmonic oscillators with which we model the microscopic 
degrees of freedom on the slab, 

\begin{equation}
 S_0[\phi] = \frac{1}{2}\int d^2x \partial_\mu\phi\partial^\mu\phi,
\end{equation}

\begin{equation}
 S_0[q_n] = \frac{1}{2} \sum_n \int dt\,  m_n \left({\dot q}_n^2 - \omega_n^2 q_n^2 \right),
\end{equation}
and the interaction term is assumed as 

\begin{equation}
 S_{\rm int}[\phi,q_n] =- \sum_n \int d^2x \, \lambda(t) \phi(t,x) q_n(t) \delta(x - \frac{L}{2}),
\end{equation} where $\lambda(t)$ is the time-dependent coupling between the 
scalar field and the microscopic degrees of freedom on the slab. With this time dependent coupling
we are modelling the time dependent conductivity of the slab, which could be produced 
by a time dependent charge carrier density within the semiconducting slab.

Our system of interest is the massless scalar field. Therefore, we will proceed integrating out the 
harmonic oscillator variables $q_n$ in order to obtain an effective description of the 
evolution of the field. This non-unitary evolution is described, mainly, by two environmentally induced 
effects: dissipation and noise. After tracing over the environment oscillators, we are able to 
write an effective action for the quantum open system  \cite{EA} of the form

\begin{equation}
 A[\phi,\phi'] = S_0[\phi] - S_0[\phi'] + \delta A[\phi, \phi'],
\end{equation}
where $\phi$ and $\phi'$ are each of the CTP branches for the field. The influence 
(Feynmann-Vernon) action is given by

\begin{eqnarray}
 \delta A[\phi, \phi'] &=& \int d^2x\int d^2x' \Delta(t,x) D(t;t') \delta(x - \frac{L}{2}) 
\delta(x' - \frac{L}{2}) \Sigma(t',x')  \nonumber \\
&+& i  \int d^2x\int d^2x' \Delta(t,x) N(t;t') \delta(x - \frac{L}{2}) 
\delta(x' - \frac{L}{2}) \Delta(t',x') 
\label{IF}\end{eqnarray}
where we have defined $\Delta(t,x) = \phi(t,x) - \phi'(t,x)$ and $\Sigma(t,x) = 1/2 [\phi(t,x) + \phi'(t,x)]$. 

The influence action is in general non-local and complex, and contains all the information 
about the environment (i.e. temperature, spectral density, dissipation rates, and noise 
properties). The kernel in the real part of $\delta A$ is the dissipation kernel $D(t;t')$, and $N(t;t')$ in the imaginary part is related with the quantum fluctuations or noise coming from the external bath. These kernels 
are usually related by means of a fluctuation-dissipation relation. The 
time-dependent coupling strenght is included in the definition of the noise and dissipation 
kernels in Eq.(\ref{IF}). On general grounds, one can expect that $D(t;t')=\lambda(t)\lambda(t') \tilde{D}(t-t')$ and 
$N(t;t')=\lambda(t)\lambda(t') \tilde{N}(t-t')$. 

Different environments can be considered. Each of them is characterized by a spectral density. 
The ohmic environment is the most studied case in the literature and produces a dissipative force that 
is proportional to the velocity. The supraohmic case, on the one hand, is generally used to 
model the interaction between defects and phonons in metals and also to mimic the interaction between a
charge and its own electromagnetic field \cite{Legget}. On the other hand, the quantum behaviour of “free” electrons in mesoscopic systems is affected by their interaction with the environment, which, for example in such cases, consist of other electrons, phonons, photons or scatterers. Which environment is more relevant for the dissipative phenomena 
generally depends on the temperature. For instance, the temperature dependence of the weak-localization correction to the
conductivity reveals in metals that electron-electron interactions dominate over the phonon contribution to
decoherence at the low temperature regime. Therefore, in order to mimic a real material for the slab in
the cavity, it will be relevant to consider spectral densities such the ones mentioned above.
Only in the particular case in which the coupling is constant $\lambda(t) = \lambda_0$, and we consider an ohmic enviroment at very 
high temperature, the noise kernel is $N(t - t') \sim \lambda_0^2 \delta(t - t')$ and the 
dissipation kernel reads $D(t - t') \sim \lambda_0^2 \dot\delta(t - t')$. Therefore, the influence 
action is local in time. It is worth to remark that, for these models, the fluctuation-dissipation relation 
reads 
\begin{equation}
 N(t) = 2k_B T \int_{-\infty}^{+\infty} ds D(s),
\end{equation}
which is the classical Einstein formula.

In the general case, the equation of motion of the scalar field can be obtained from the effective action $A[\phi,\phi']$ 
by $\frac{\delta A}{\delta \phi}\vert_{\phi = \phi'} = 0$: 

\begin{equation}
 \Box \phi + \delta(x - \frac{L}{2})\int_0^t ds D(t;s) \phi(\frac{L}{2}, s) = 0.
\end{equation}
This equation corresponds to an average over all the possible noise  realizations. In order to 
extract the noise information from the effective action, it is possible to write down the 
imaginary part of the influence action in terms of a stochastic noise force ${\cal F}$, which 
is coupled to the main system of interest, and it is defined by means off a Gaussian probability 
distribution $P[ {\cal F}]$ like

\begin{equation}
P[{\cal F}] = N_{{\cal F}} \exp\left\{-\frac{1}{2} \int_0^t ds \int_0^s ds' {\cal F}(s)N(s;s')^{-1}
{\cal F}(s')\right\},
\end{equation}
where $N_{{\cal F}}$ is a normalization constant. Therefore, the imaginary part of $\delta A$ can be 
written in terms of the stochastic force source as

\begin{eqnarray}
 \int {\cal D}{\cal F}(t) &P[{\cal F}]& \exp\left\{-\frac{i}{\hbar} \Delta (t,\frac{L}{2}) {\cal F}(t)\right\} \nonumber \\
&=& \exp\left\{\frac{i}{\hbar}\int_0^t ds \int_0^s ds' \Delta(s,\frac{L}{2}) N(s;s') \Delta(s',\frac{L}{2})\right\}.\label{noisesource}
\end{eqnarray}

Finally, from Eq.(\ref{noisesource}) we can derive an equation of motion for the massless scalar field that 
takes into account the noise. This is a Langevin-like equation, where the full dynamics of the 
field is also determined by the presence of the stochasticity induced by the environment via ${\cal F}$. The 
Langevin equation now reads as

\begin{equation}
 \Box \phi + \delta(x - \frac{L}{2})\int_0^t ds D(t;s) \phi(\frac{L}{2}, s) = {\cal F}(t)\delta(x - \frac{L}{2}).
\end{equation}

The stochastic noise ${\cal F}$ is characterised by the probability distribution $P[{\cal F}]$, and also 
by the correlation functions

$$\langle {\cal F}(t)\rangle = 0 ~~~~~;~~~~~ \langle {\cal F}(t){\cal F}(t') \rangle = N(t;t')\,\, .$$

\section{The dynamics of the resonant mode}

The aim of this section is to obtain a dynamical equation for the
evolution of a
single mode of the quantum field, starting from the effective equation
derived in the previous section.
We have already seen that, in the absence of dissipation, if the
conductivity of the slab
has periodic changes with frequency $\Omega$, only the resonant mode with
$\tilde k^0=\frac{1}{2}\Omega$ is relevant at long times. We will
now assume that this is true even in
the presence of dissipation. Although reasonable, this is an
unjustified assumption, and we hope to further analyze it in a
forthcoming publication. Even with this simplification, we will see
that the derivation of the dynamical equation of the mode
is a non-trivial task.

Let us write the dissipation kernel as
\begin{equation}
 D(t;s)=V_0\delta(t-s)+ d(t;s)\, .
\label{perturb}
\end{equation}
In
the particular case $d(t;s)=(V(t)-V_0)\delta(t-s)$,
with $V(t)$ a periodic function of frequency $\Omega$ 
we recover the model of Section 2. 
In that case,
we have shown that even when
$V(t)$ differs several orders of magnitude from $V_0$,
the oscillations in the wavenumber have a relative amplitude
$\epsilon_n\ll 1$ (see the discussion after Eq.(\ref{epsilon})).
Therefore, we will assume
that the main contribution to the wavenumber comes from
the term 
proportional to $V_0$, and we will treat the 
contributions of both $d(t;s)$ and the noise kernel as small
corrections. It is worth to emphasize that we are not assuming
that $D(t,s)\approx V_0\delta(t-s)$, but that the effect of $d(t;s)$
and the noise
on the instantaneous wavenumber $k(t)$ can be treated
perturbatively.

The resonant mode of the field can be written as
\begin{equation}
 u(x,t)=P(t) \Psi(x,t)
\label{resmode}
\end{equation}
where $\Psi(x,t)$ is given in Eq.(\ref{sol2}). In order to simplify the notation,
from now on we omit the subindex $m$, since we are considering a single mode. 
Inserting this particular mode
into the Langevin equation for the field,
we find that the boundary condition at the position of the slab becomes
\begin{equation}
 2 P(t)k(t)\cos[\frac{k(t)L}{2}]=-\int_0^t ds\, 
D(t,s)P(s)\sin[\frac{k(t)L}{2}] + \sqrt{\frac{L}{2}} {\cal F}(t)\, .
\label{bc}
\end{equation}
It is worth to stress that, when the dissipation kernel is local and in the absence of
noise, the
amplitude of the mode $P(t)$ factorizes
and the boundary condition fixes the value of the time dependent
wavenumber $k(t)$. In the general case,
$k(t)$ becomes a nonlocal function of $P(t)$. 
To see this explicitly,
we solve Eq.(\ref{bc}) assuming that $d(t;s)$ and ${\cal F}(t)$
induce small time dependent corrections to $k(t)$.
We write $k(t)=k^0+\Delta k(t)$, where $k^0$ is the
solution of the unperturbed (static) problem, i.e.
\begin{equation}
2k^0=- V_0\tan[\frac{k^0L}{2}]\, .
\end{equation}
To first order in $d(t;s)$ and ${\cal F}$ we have
\begin{equation}
 \Delta k(t)P(t)=\frac{k^0}{V_0[1+\frac{V_0 L}{4}+\frac{(k^0)^2L}{V_0}]}\left[\int_0^tds\, d(t;s)P(s)- 
\tilde{\cal F}(t)\right]
\label{pertsol}
\end{equation}
where $\tilde{\cal F}=\sqrt{L/2} \, {\rm cosec}(k^0L/2)$.
As anticipated, the nonlocal boundary condition becomes a nonlocal
relation between $\Delta k(t)$ and $P(t)$.

The next step is to insert the field mode into the Klein Gordon
equation of the field, at both sides of the slab. Of course
the single mode will not be an exact solution of the Klein Gordon
equation, which would
involve the coupling to the other modes. Therefore, in order to get
the dynamical equation for $P(t)$, we impose
$(\Psi(x,t),\Box u)=0$. After some simple calculations we get,
\begin{equation}
 \ddot P + ((k^0)^2 +2k^0\Delta k(t)) P + g^{(A)}(2\dot P\dot{\Delta k}+P\ddot{\Delta k})=0
\label{eqlangmode}
\end{equation}
where $g^{(A)}$ corresponds to  $g^{(A)}_{mm}$ is defined in Section 2, and can be evaluated to
lowest order, setting $d(t;s)=0$ and ${\cal F}=0$. Note that
in this approximation this coupling constant is independent of time.

Let us analyze the resulting equation for the field mode. To begin with, it is important to note
that the term proportional to $\Delta k(t)P(t)$
contains nonlocal dissipation an additive noise. Moreover, if we
write $d(t;s)= (V(t)-V_0)\delta (t-s) + \tilde d(t;s)$
the local term corresponds to a time dependent frequency. Depending on the details 
of the environment, we
expect that, at high temperatures,
$\tilde d(t-s)\sim \dot\delta (t-s)$ so the usual dissipative term
proportional to $\dot P$ is recovered. Moreover, in this limit
the fluctuation-dissipation theorem implies a white noise. In any
other case, we expect nonlocal dissipation and colored noise.

In the absence of dissipation, the last term in the lhs of
Eq.(\ref{eqlangmode}), proportional to $g^{(A)}$,
can be neglected when the external frequency satisfies $\Omega = 2
\tilde k^0$. However, in general this is not the case.
Noting that $2\dot P\dot \Delta k+P\ddot{\Delta k} = \ddot {(P{\Delta k})} - \ddot P
\Delta k$, it is simple to find an expression
for this additional term by taking two derivatives of
Eq.(\ref{pertsol}). The result is a rather complicated expression:
\begin{eqnarray}
&& 2\dot P\dot \Delta k+P\ddot{\Delta k} = - \ddot P
\Delta k + \frac{k^0}{V_0[1+\frac{V_0 L}{4}+\frac{(k^0)^2L}{V_0}]}\left[
P(\dot d(t;t)+ \partial_t d(t;t))\right .\nonumber\\
&+& \left . d(t;t)\dot P +
\int_0^tds\, \partial_t^2 d(t;s)P(s)- 
\ddot{\tilde{\cal F}}(t)\right]\,\, .
\end{eqnarray}
The main
consequence is that, after inserting this result into Eq.(\ref{eqlangmode}),
the final equation for $P(t)$ can be written as
 \begin{equation}
 \ddot P + \omega^2_{\rm eff}(t) P + \int_0^tds\,  \, d_{\rm eff}(t,s) P(s) = {\cal F}_{\rm eff}(t)\, .
\label{eqlangmodefin}
\end{equation}
In summary, we have shown that the amplitude of the mode satisfies a
Langevin equation with nonlocal dissipation and colored
noise. The dissipation ($d_{\rm eff}$) and noise (${\cal F}_{\rm eff}$) kernels are related to the ones in
the Langevin equation for the field derived in Section 3,
although they are not exactly the same because of the presence of the
term proportional to $g^{(A)}$.
All the information about dissipation and noise comes from the
nonlocal boundary condition that the field satisfies
on the slab.

\section{Conclusions}

In this paper we have argued that the natural approach to analyze the dynamical Casimir effect is that of the quantum open systems. 
The coupling between the degrees of freedom of the vacuum field and those of the (imperfect) mirrors generate an
infuence functional for the vacuum field that contains the information about dissipation and noise. The dynamics of the 
field is described by a Langevin equation with dissipative and noise kernels concentrated on the position of the mirror.

When the system is under the influence of periodic, time dependent external conditions, it is possible to have parametric amplification
in some of the modes of the field. If the spectrum of the unperturbed system is not equidistant, one expects the amplification to
occur for a single mode. Assuming that this is the case even in the presence of dissipation, we have shown that the dynamics of
the mode can also be described by a Langevin equation with nonlocal dissipation and colored noise. This equation is similar to the one derived in the context of quantum brownian motion, but now the frequency and the (nonlocal) kernels have a periodic time dependence. 

We have considered a number of simplifications in order to illustrate the above points: we worked with a scalar field in $1+1$
dimensions, and we {\it assumed} that the parametric amplification can be described in terms of a single mode. The last 
assumption 
deserve further investigation. Moreover, we have not attempted to compute the influence of dissipation and noise 
on  the number of photons created. This seems to be a rather difficult task, because of the complexity of the Langevin equation for
the mode. Some additional simplifications could be necessary in order to estimate these effects (along the lines
of Ref.\cite{Dodonovdis}, for instance). However we think that we have clarified, from  a conceptual point of view, the origin of
the description of the DCE in terms of a noisy and damped oscillator.

This work has been supported by  Universidad de Buenos Aires,
CONICET and ANPCyT.

\end{document}